\documentclass[onecolumn,twoside]{IEEEtran}
\usepackage{mathpazo}
\usepackage{times}
\usepackage{color}
\usepackage{amsmath}
\usepackage{amsfonts}
\usepackage{latexsym}
\usepackage{amssymb,amsbsy}
\usepackage{cite,url}
\usepackage{mathtools}
\usepackage{amscd}

\usepackage{xcolor,stackengine}

\usepackage{upref}
\usepackage{theorem}
\usepackage{graphicx}
\usepackage{psfrag}
\usepackage{multirow}

\theoremstyle{plain}
\theorembodyfont{\normalfont\slshape}

\newtheorem{thm}{Theorem$\!$}
\newenvironment{theorem}
{\begin{thm}\hspace*{-1ex}{\bf.}}{\end{thm}}

\newtheorem{lem}[thm]{Lemma$\!$}
\newenvironment{lemma}{\begin{lem}\hspace*{-1ex}{\bf.}}{\end{lem}}

\newtheorem{prop}[thm]{Proposition$\!$}

\newtheorem{cor}[thm]{Corollary$\!$}
\newenvironment{corollary}{\begin{cor}\hspace*{-1ex}{\bf.}}{\end{cor}}

\newtheorem{defn}[thm]{Definition$\!$}
\newenvironment{definition}{\begin{defn}\hspace*{-1ex}{\bf.}}{\end{defn}}

\newtheorem{xmpl}[thm]{Example$\!$}
\newenvironment{example}{\begin{xmpl}\hspace*{-1ex}{\bf.}}{\hfill$\Box$\end{xmpl}}

\newtheorem{cnstr}{Construction$\!$}

\newenvironment{construction}{\begin{cnstr}\hspace*{-1ex}{\bf.}}{\end{cnstr}}

\newcounter{enumrom}
\renewcommand{\theenumrom}{(\roman{enumrom})}

\makeatletter
\renewcommand{\@endtheorem}{\endtrivlist}
\makeatother

\makeatletter
\renewcommand{\thefigure}{{\@arabic\c@figure}}
\renewcommand{\fnum@figure}{{\bf Figure\,\thefigure}}
\makeatother

\newcommand{\cE}{\mathcal{E}}
\newcommand{\cF}{\mathcal{F}}
\newcommand{\cG}{\mathcal{G}}

\newcommand{\cT}{\mathcal{T}}

\newcommand{\cV}{\mathcal{V}}

\newcommand{\mathset}[1]{\left\{#1\right\}}
\newcommand{\abs}[1]{\left|#1\right|}
\newcommand{\ceilenv}[1]{\left\lceil #1 \right\rceil}
\newcommand{\floorenv}[1]{\left\lfloor #1 \right\rfloor}
\newcommand{\parenv}[1]{\left( #1 \right)}

\newcommand{\bracenv}[1]{\left\{ #1 \right\}}

\newcommand{\be}[1]{\begin{equation}\label{#1}}
\newcommand{\ee}{\end{equation}}

\renewcommand{\le}{\leqslant}
\renewcommand{\leq}{\leqslant}

\renewcommand{\geq}{\geqslant}

\renewcommand{\Bbb}{\mathbb}

\newcommand{\Cref}[1]{Co\-ro\-lla\-ry\,\ref{#1}}

\renewcommand{\Bbb}{\mathbb}

\newcommand{\N}{{\Bbb N}}

\newcommand{\Z}{{\Bbb Z}}

\DeclareMathOperator{\wt}{wt}
\DeclareMathOperator{\suff}{Suff}
\DeclareMathOperator{\pref}{Pref}
\DeclareMathOperator{\rll}{RLL}
\newcommand{\stan}{S} 
\newcommand{\ttan}{T} 
\newcommand{\cttan}{\cT} 
\DeclareMathOperator*{\xder}{\Longrightarrow}

\DeclareMathOperator*{\xdedup}{\Longleftarrow}

\newcommand{\der}{\xder\limits}

\newcommand{\dedup}{\xdedup\limits}

\DeclareMathOperator{\irr}{Irr}
\newcommand{\ew}{\epsilon}
\newcommand{\ccap}{\mathsf{cap}}

\outer\def\proclaim #1. #2\par{\medbreak
 \noindent{\bf#1.\enspace}{\sl#2\par}%
 \ifdim\lastskip<\medskipamount \removelastskip\penalty55\medskip\fi}

\begin{document}


\title{\textbf{Duplication-Correcting Codes for Data Storage\\ in the DNA of Living Organisms}}

\author{\large
  Siddharth Jain,
  Farzad Farnoud (Hassanzadeh),~\IEEEmembership{Member,~IEEE},
  Moshe Schwartz,~\IEEEmembership{Senior Member,~IEEE},
  Jehoshua Bruck,~\IEEEmembership{Fellow,~IEEE}%
  \thanks{The material in this will be presented in part at the 2016 IEEE International Symposium on Information Theory.}%
  \thanks{Siddharth Jain is with the Department
    of Electrical Engineering, California Institute of Technology, Pasadena, CA 91125, USA (e-mail: sidjain@caltech.edu).}%
  \thanks{Farzad Farnoud (Hassanzadeh) is with the Department
    of Electrical Engineering, California Institute of Technology, Pasadena, CA 91125, USA (e-mail: farnoud@caltech.edu).}%
  \thanks{Moshe Schwartz is with the Department
    of Electrical and Computer Engineering, Ben-Gurion University of the Negev,
    Beer Sheva 8410501, Israel
    (e-mail: schwartz@ee.bgu.ac.il).}%
  \thanks{Jehoshua Bruck is with the Department
    of Electrical Engineering, California Institute of Technology, Pasadena, CA 91125, USA (e-mail: bruck@paradise.caltech.edu).}%
  \thanks{This work was supported in part by the NSF Expeditions in Computing Program (The Molecular Programming Project).}
}

\maketitle

\begin{abstract}
The ability to store data in the DNA of a living organism has
applications in a variety of areas including synthetic biology and
watermarking of patented genetically-modified organisms. Data stored
in this medium is subject to errors arising from various mutations,
such as point mutations, indels, and tandem duplication, which need to
be corrected to maintain data integrity. In this paper, we provide
error-correcting codes for errors caused by tandem duplications, which
create a copy of a block of the sequence and insert it in a tandem
manner, i.e., next to the original. In particular, we present two
families of codes for correcting errors due to tandem-duplications of
a fixed length; the first family can correct any number of errors
while the second corrects a bounded number of errors. We also study
codes for correcting tandem duplications of length up to a given
constant $k$, where we are primarily focused on the cases of $k=2,3$.
Finally, we provide a full classification of the sets of lengths
allowed in tandem duplication that result in a unique root for all
sequences.
\end{abstract}


\section{Introduction}
\IEEEPARstart{D}{ata} storage in the DNA of living organisms
(henceforth \emph{live DNA}) has a multitude of applications. It can
enable in-vivo synthetic-biology methods and algorithms that need
``memory,'' e.g., to store information about their state or record
changes in the environment. Embedding data in live DNA also allows
watermarking genetically-modified organisms (GMOs) to verify
authenticity and to track unauthorized
use~\cite{AriOha04,HeiBar07,Micetal12}, as well as labeling organisms
in biological studies~\cite{WonWonFoo03}. DNA watermarking can also be
used to tag infectious agents used in research laboratories to
identify sources of potential malicious use or accidental
release~\cite{Jupetal10}. Furthermore, live DNA can serve as a
protected medium for storing large amounts of data in a compact format
for long periods of time~\cite{WonWonFoo03,Bal13}. An additional
advantage of using DNA as a medium is that data can be disguised as
part of the organisms original DNA, thus providing a layer of
secrecy~\cite{CleRisBan99}.

While the host organism provides a level of protection to the
data-carrying DNA molecules as well as a method for replication, the
integrity of the stored information suffers from mutations such as
tandem duplications, point mutations, insertions, and
deletions. Furthermore, since each DNA replication may introduce new
mutations, the number of such deleterious events increases with the
number of generations. As a result, to ensure decodability of the
stored information, the coding/decoding scheme must be capable of a
level of error correction. Motivated by this problem, we study
designing codes that can correct errors arising from tandem
duplications. In addition to improving the reliability of data storage
in live DNA, studying such codes may help to acquire a better
understanding of how DNA stores and protects biological information in
nature.

Tandem duplication is the process of inserting a copy of a segment of
the DNA adjacent to its original position, resulting in a \emph{tandem
  repeat}. A process that may lead to a tandem duplication is
\emph{slipped-strand mispairings}~\cite{MunHel04} during DNA
replication, where one strand in a DNA duplex is displaced and
misaligned with the other. Tandem repeats constitute about 3\% of the
human genome~\cite{Lanetal01} and may cause important phenomena such
as chromosome fragility, expansion diseases, silencing
genes~\cite{Usd08}, and rapid morphological variation~\cite{FonGar04}.

Different approaches to the problem of error-control for data stored
in live DNA have been proposed in the literature. In the work of Arita
and Ohashi~\cite{AriOha04}, each group of five bits of information is
followed by one parity bit for error detection. Heider and
Barnekow~\cite{HeiBar07} use the extended $[8,4,4]$ binary Hamming
code or repetition coding to protect the data. Yachie et
al.~\cite{YacOhaYosTom08} propose to enhance reliability by inserting
multiple copies of the data into multiple regions of the genome of the
host organism. Finally, Haughton and Balado~\cite{HauBal13} present an
encoding method satisfying certain biological constraints, which is
studied in a substitution-mutation model. None of the aforementioned
encodings, with the possible exception of repetition coding, are
designed to combat tandem duplications, which is the focus of this
paper. While repetition coding can correct duplication errors, it is
not an efficient method because of its high redundancy.

It should also be noted that error control for storage in live DNA is
inherently different from that in DNA that is stored outside of a
living organism (see~\cite{Yazdi2015} for an overview), since the
latter is not concerned with errors arising during organic DNA
replication.

In this work, we ignore the potential biological effects of embedding
data into the DNA. Furthermore, constructing codes that, in addition
to tandem-duplication errors, can combat other types of errors, such
as substitutions, are postponed to a future work.

We also note that tandem duplication, as well as other duplication
mechanisms, were studied in the context of information theory
\cite{FarSchBru15,JaiFarBru15,FarSchBru16}. However,
these works used duplications as a generative process, and attempted
to measure its capacity and diversity.  In contrast, we consider
duplications as a noise source, and design error-correcting codes to
combat it.

We will first consider the tandem-duplication channel with
duplications of a fixed length $k$. For example with $k=3$, after a
tandem duplication, the sequence $ACAGT$ may become
$ACAG\underline{CAG}T$, which may further become
$ACA\underline{ACA}GCAGT$ where the copy is underlined. In our
analysis, we provide a mapping in which tandem duplications of length
$k$ are equivalent to insertion of $k$ zeros. Using this mapping, we
demonstrate the strong connection between codes that correct
duplications of a fixed length and Run-Length Limited (RLL)
systems. We present constructions for codes that can correct an
unbounded number of tandem duplications of a fixed length and show
that our construction is optimal, i.e., of largest size.  A similar
idea was used in \cite{DolAna10}, where codes were constructed for
duplication-error correction with the number of tandem duplications
restricted to a given size $r$ and a duplication length of $1$ only. In
this paper, we generalize their result by constructing optimal (i.e.,
maximum size) error-correcting codes for arbitrary duplication length
$k$ and with no restriction on the number of tandem duplications.

We then turn our attention to codes that correct $t$ tandem
duplications (as opposed to an unbounded number of duplications), and
show that these codes are closely related to constant-weight codes in
the $\ell_1$ metric.

We also consider codes for correcting duplications of bounded
length. Here, our focus will be on duplication errors of length at
most $2$ or $3$, for which we will present a construction that
corrects any number of such errors. In the case of duplication length
at most $2$ the codes we present are optimal.

Finally, when a sequence has been corrupted by a tandem-duplication
channel, the challenge arises in finding the \emph{root} sequences
from which the corrupted sequence could be generated. A root sequence
cannot be the result of tandem-duplication mutation on some other
sequence. For example, for the sequence $ACGT\underline{GT}$, with
$GTGT$ as a tandem-duplication error, a root sequence would be $ACGT$
since $ACGTGT$ can be generated from $ACGT$ by doing a tandem
duplication of length $2$ on $GT$. But there can be sequences that
have more than one root. For example, the sequence $ACGCACGCG$ can be
generated from $ACG$ by doing a tandem duplication of $CG$ first,
followed by a tandem duplication of $ACGC$. Alternatively, it can also
be generated from $ACGCACG$ by doing a tandem duplication of the
suffix $CG$. Hence, $ACGCACGCG$ has two roots. However, if we restrict
the length of duplication to $2$ in the previous example, then
$ACGCACGCG$ has only one root i.e., $ACGCACG$. This means that the
number of roots that a sequence can have depends on the set of
duplication lengths that are allowed, and the size of the alphabet.
We provide in Section \ref{sec:roots} a complete classification of the
parameters required for the unique-root property.  This unique-root
property for fixed length, $2$-bounded and $3$-bounded
tandem-duplication channels allows us to construct error-correcting
codes for them.

The paper is organized as follows. The preliminaries and notation are
described in Section~\ref{sec:prel}. In Sections~\ref{sec:k-tandem}
and~\ref{sec:le-k-tandem} we present the results concerning
duplications of a fixed length $k$ and duplications of length at most
$k$, respectively. In Section \ref{sec:roots}, we fully classify
tandem-duplication channels which have a unique root. We conclude with
some open questions in Section \ref{sec:conc}.

\section{Preliminaries}\label{sec:prel}

We let $\Sigma$ denote some finite alphabet, and $\Sigma^*$ denote the
set of all finite strings (words) over $\Sigma$. The unique empty word
is denoted by $\epsilon$. The set of finite non-empty words is denoted
by $\Sigma^+=\Sigma^*\setminus\mathset{\epsilon}$. Given two words
$x,y\in\Sigma^*$, their concatenation is denoted by $xy$, and $x^t$
denotes the concatenation of $t$ copies of $x$, where $t$ is some
positive integer. By convention, $x^0=\epsilon$. We normally index the
letters of a word starting with $1$, i.e., $x=x_1 x_2\dots x_{n}$,
with $x_i\in \Sigma$. With this notation, the $t$-prefix and
$t$-suffix of $x$ are defined by
\begin{align*}
  \pref_t(x) &= x_1 x_2 \dots x_t, \\
  \suff_t(x) &= x_{n-t+1} x_{n-t+2} \dots x_n.
\end{align*}

Given a string $x\in\Sigma^*$, a \emph{tandem duplication of length
  $k$} is a process by which a contiguous substring of $x$ of length
$k$ is copied next to itself. More precisely, we define the
tandem-duplication rules, $\ttan_{i,k}:\Sigma^*\to\Sigma^*$, as
\[
\ttan_{i,k}(x)=\begin{cases}
uvvw & \text{if $x=uvw$, $\abs{u}=i$, $\abs{v}=k$}\\
x & \text{otherwise.}
\end{cases}
\]
Two specific sets of duplication rules would be of interest to us
throughout the paper.
\begin{align*}
\cttan_{k} & =\mathset{\left.\ttan_{i,k}~\right|~i\geq0},\\
\cttan_{\leq k} & =\mathset{\left.\ttan_{i,k'}~\right|~i\geq0, 1\leq k'\leq k}.
\end{align*}

Given $x,y\in\Sigma^*$, if there exist $i$ and $k$ such that
\[ y=\ttan_{i,k}(x),\]
then we say $y$ is a direct descendant of $x$, and denote it
by
\[ x \der_k y.\]
If a sequence of $t$ tandem duplications of length $k$ is employed to
reach $y$ from $x$ we say $y$ is a $t$-descendant of $x$ and denote it
by
\[ x \der_k^t y.\]
More precisely, we require the existence of $t$ non-negative integers
$i_1,i_2,\dots,i_t$, such that
\[ y = \ttan_{i_t,k}(\ttan_{i_{t-1},k}( \dots \ttan_{i_1,k}(x)\dots )).\]
Finally, if there exists a finite sequence of tandem duplications of
length $k$ transforming $x$ into $y$, we say $y$ is a descendant of
$x$ and denote it by
\[ x \der_k^* y.\]
We note that $x$ is its own descendant via an empty sequence of tandem
duplications.

\begin{example}
Let $\Sigma=\bracenv{0,1,2,3}$ and $x=02123$. Since,
$T_{1,2}(x)=0212123$ and $T_{0,2}(0212123)=020212123$, the following hold
\begin{align*}
02123&\der_2 0212123,&02123&\der_2^2 020212123,
\end{align*}
where in both expressions, the relation could be replaced with~$\der_2^*$.
\end{example}

We define the \emph{descendant cone} of $x$ as
\[ D_k^*(x) = \mathset{ y\in\Sigma^* ~\left|~ x\der_k^* y \right. }.\]
In a similar fashion we define the \emph{$t$-descendant cone}
$D_k^t(x)$ by replacing $\der_k^*$ with $\der_k^t$ in the
definition of $D_k^*(x)$.

The set of definitions given thus far was focused on
tandem-duplication rules of substrings of length exactly $k$, i.e.,
for rules from $\cttan_k$. These definitions as well as others in this section are extended in the natural way for
tandem-duplication rules of length up to $k$, i.e., $\cttan_{\leq
  k}$. We denote these extensions by replacing the $k$ subscript with
the $\leq k$ subscript. Thus, we also have $D_{\leq k}^*(x)$ and $D_{\leq
  k}^t(x)$.

\begin{example}
Consider $\Sigma=\bracenv{0,1}$ and $x=01$. It is not difficult to see that 
\begin{align*}
D_{1}^{2}(x) & =\mathset{ 0001,0011,0111} ,\\
D_{1}^{*}(x) & =\mathset{ \left. 0^{i}1^{j} ~\right|~ i,j\in\mathbb{N}} ,\\
D_{2}^{*}(x) & =\mathset{ \left. (01)^{i} ~\right|~ i\in\mathbb{N}} ,\\
D_{\le2}^{*}(x) & =\mathset{ 0s1 ~|~ s\in\Sigma^{*}}.
\end{align*}
\end{example}

Using the notation $D_k^*$, we restate the definition of the \emph{tandem
  string-duplication system} given in~\cite{FarSchBru16}. Given a
finite alphabet $\Sigma$, a seed string $s\in\Sigma^*$, the tandem
string-duplication system is given by
\[ \stan_k = S(\Sigma,s,\cttan_k) = D_k^*(s),\]
i.e., it is the set of all the descendants of $s$ under tandem
duplication of length $k$.

The process of tandem duplication can be naturally reversed. Given a
string $y\in\Sigma^*$, for any positive integer, $t>0$, we define the
\emph{$t$-ancestor cone} as
\[ D_k^{-t}(y) = \mathset{ x\in\Sigma^* ~\left|~ x\der_k^t y\right. },\]
or in other words, the set of all words for which $y$ is a
$t$-descendant.

Yet another way of viewing the $t$-ancestor cone is by defining the
\emph{tandem-deduplication rules},
$\ttan^{-1}_{i,k}:\Sigma^*\to\Sigma^*$, as
\[
\ttan^{-1}_{i,k}(y)=\begin{cases}
uvw & \text{if $y=uvvw$, $\abs{u}=i$, $\abs{v}=k$}\\
\ew & \text{otherwise,}
\end{cases}
\]
where we recall $\ew$ denotes the empty word.  This operation takes an
adjacently-repeated substring of length $k$, and removes one of its
copies. Thus, a string $x$ is in the $t$-ancestor cone of $y$ (where
we assume $x,y\neq \epsilon$ to avoid trivialities) iff there is a
sequence of of $t$ non-negative integers $i_1,i_2,\dots,i_t$, such
that
\[ x = \ttan^{-1}_{i_t,k}(\ttan^{-1}_{i_{t-1},k}( \dots \ttan^{-1}_{i_1,k}(y)\dots )).\]
In a similar fashion we define the \emph{ancestor cone} of $y$ as
\[ D^{-*}_k(y) = \mathset{ x\in\Sigma^* ~\left|~ x\der_k^* y \right. }.\]
By flipping the direction of the derivation arrow, we let $\dedup$ denote
deduplication. Thus, if $y$ may be deduplicated to obtain $x$ in a single
step we write
\[y\dedup_k x.\]
For multiple steps we add $*$ in superscript.
\begin{example}
We have
\begin{align*}
 0212123&\dedup_2 02123,&020212123&\dedup_2^2 02123,
\end{align*}
and
\[D^{-*}_2(020212123) =\mathset{020212123,0212123,0202123,02123}.\]
\end{example}

A word $y\in\Sigma^*$ is said to be \emph{irreducible} if there is nothing
to deduplicate in it, i.e., $y$ is its only ancestor, meaning
\[ D^{-*}_k(y) = \mathset{y}.\]
The set of irreducible words is denoted by $\irr_k$. We will find it useful
to denote the set of irreducible words of length $n$ by
\[\irr_k(n) = \irr_k \cap \Sigma^n.\]
The ancestors of $y\in\Sigma^*$ that cannot be further deduplicated,
are called the \emph{roots} of $y$, and are denoted by
\[ R_k(y) = D^{-*}_k(y) \cap \irr_k.\]

Note that since the aforementioned definitions extend to tandem-duplication rules of length up to $k$, we also have $\stan_{\leq k}$,
$D^{-t}_{\leq k}(y)$, $D^{-*}_{\leq k}(y)$, $\irr_{\leq k}$,
$\irr_{\leq k}(n)$, and $R_{\leq k}(y)$. In some previous works (e.g.,
\cite{LeuMarMit05}), $\stan_k$ is called the
\emph{uniform-bounded-duplication system}, whereas $\stan_{\leq k}$ is
called the \emph{bounded-duplication system}.
\begin{example}
For the binary alphabet $\Sigma=\bracenv{0,1}$,
\begin{equation*}
\irr_{\le2}=\bracenv{0,1,01,10,010,101},
\end{equation*}
and for any alphabet that contains $\bracenv{0,1,2,3}$,
\begin{align*}
R_2(020212123)&=\bracenv{02123},\\
R_{\le4}(012101212)&=\bracenv{012,0121012}.
\end{align*}
\end{example}

Inspired by the DNA-storage scenario, we now define error-correcting
codes for tandem string-duplication systems.

\begin{definition}
  An $(n,M;t)_k$ code $C$ for the $k$-tandem-duplication channel is a
  subset $C\in\Sigma^n$ of size $\abs{C}=M$, such that for each
  $x,y\in C$, $x\neq y$,
  \[D_k^t(x)\cap D_k^t(y) = \emptyset.\]
  Here $t$ stands for either a non-negative integer, or $*$. In the
  former case we say the code can correct $t$ errors, whereas in the
  latter case we say the code can correct all errors. In a similar
  fashion, we define an $(n,M;t)_{\leq k}$ by replacing all ``$k$''
  subscripts by ``$\leq k$''.
\end{definition}

Assume the size of the finite alphabet is $\abs{\Sigma}=q$. We then
denote the size of the largest $(n,M;t)_k$ code over $\Sigma$ by
$A_q(n;t)_k$. The capacity of the channel is then defined as
\[ \ccap_q(t)_k = \limsup_{n\to\infty} \frac{1}{n}\log_q A_q(n;t)_k. \]
Analogous definitions are obtained by replacing $k$ with $\leq k$ or
by replacing $t$ with $*$.

\section{$k$-Tandem-Duplication Codes}\label{sec:k-tandem}

In this section we consider tandem string-duplication systems where
the substring being duplicated is of a constant length $k$. Such
systems were studied in the context of formal
languages~\cite{LeuMarMit05} (also called
\emph{uniform-bounded-duplication systems}), and also in the context
of coding and information theory~\cite{FarSchBru16}.

In~\cite{LeuMarMit05} it was shown that for any finite alphabet $\Sigma$,
and any word $x\in\Sigma^*$, under $k$-tandem duplication $x$ has a unique
root, i.e.,
\[\abs{R_k(x)}=1.\]
Additionally, finding the unique root may be done efficiently, even by
a greedy algorithm which searches for occurrences of $ww$ as
substrings of $x$, with $\abs{w}=k$, removing one copy of $w$, and
repeating the process. This was later extended in~\cite{Leu07}, where
it was shown that the roots of a regular languages also form a regular
language. In what follows we give an alternative elementary proof to
the uniqueness of the root. This proof will enable us to easily
construct codes for $k$-tandem-duplication systems, as well as to
state bounds on their parameters. The proof technique may be seen as
an extension of the string-derivative technique used in
\cite{DolAna10}, which was applied only for $k=1$ over a binary
alphabet.

We also mention~\cite{FarSchBru16}, in which $\stan_k$
was studied from a coding and information-theoretic perspective. It
was shown there that the capacity of all such systems is $0$. This
fact will turn out to be extremely beneficial when devising
error-correcting codes for $k$-tandem-duplication systems.

Throughout this section, without loss of generality, we assume
$\Sigma=\Z_q$. We also use $\Z_q^*$ to denote the set of all finite
strings of $\Z_q$ (not to be confused with the non-zero elements of
$\Z_q$), and $\Z_q^{\geq k}$ to denote the set of all finite strings
over $\Z_q$ of length $k$ or more.

We shall require the following mapping, $\phi_k:\Z_q^{\geq k}\to \Z_q^k\times
\Z_q^*$.  The mapping is defined
by,
\[ \phi_k(x) = ( \pref_k(x), \suff_{\abs{x}-k}(x)-\pref_{\abs{x}-k}(x) ),\]
where subtraction is performed entry-wise over $\Z_q$. We easily
observe that $\phi_k$ is a bijection between $\Z_q^n$ and
$\Z_q^k\times \Z_q^{n-k}$ by noting that we can recover $x$ from
$\phi_k(x)$ in the following manner: first set $x_i=\phi_k(x)_i$, for
all $1\leq i\leq k$, and for $i=k+1,k+2,\dots$, set
$x_i=x_{i-k}+\phi_k(x)_i$, where $\phi_k(x)_i$ denotes the $i$th
symbol of $\phi_k(x)$. Thus, $\phi_k^{-1}$ is well defined.

Another mapping we define is one that injects $k$ consecutive zeros
into a string. More precisely, we define $\zeta_{i,k} :
\Z_q^k\times\Z_q^* \to \Z_q^k\times \Z_q^*$, where
\[
\zeta_{i,k}(x,y)=\begin{cases}
(x,u0^k w) & \text{if $y=uw$, $\abs{u}=i$}\\
(x,y) & \text{otherwise.}
\end{cases}
\]

The following lemma will form the basis for the proofs to follow.
\begin{lemma}
  \label{lem:commute}
  The following diagram commutes:
  \[
  \begin{CD}
    \Z_q^{\geq k}   @>\ttan_{i,k}>>    \Z_q^{\geq k} \\
    @VV\phi_k V                    @VV\phi_k V \\
    \Z_q^k\times \Z_q^*    @>\zeta_{i,k}>>  \Z_q^k\times\Z_q^*
  \end{CD}
  \]
  i.e., for every string $x\in\Z_q^{\geq k}$,
  \[ \phi_k(\ttan_{i,k}(x)) = \zeta_{i,k}(\phi_k(x)).\]
\end{lemma}

Before presenting the proof, we provide an example for the diagram of
the lemma.

\begin{example}\label{ex:diagram}
  Assume $\Sigma=\Z_4$. Starting with $02123$ and letting $i=1$ and
  $k=2$ leads to
  \[
  \begin{CD}
    02123   @>\ttan_{1,2} >>    021\underline{21}23\\
    @VV\phi_2 V                    @VV\phi_2 V \\
    (02,102)    @>\zeta_{1,2}>>  (02,1\underline{00}02)
  \end{CD}
  \]
  where the inserted elements are underlined.
 \end{example}
\begin{IEEEproof}
  Let $x\in\Z_q^{\geq k}$ be some string, $x=x_1 x_2 \dots
  x_n$. Additionally, let $\phi_k(x)=(y,z)$ with $y=y_1 \dots y_k$,
  and $z=z_1 \dots z_{n-k}$. We first consider the degenerate case, where
  $i\geq n-k+1$. In that case, $T_{i,k}(x)=x$, and then by definition
  $\zeta_{i,k}(y,z)=(y,z)$ since $z$ does not have a prefix of length at least
  $n-k+1$. Thus, for $i\geq n-k+1$ we indeed have
  \[ \phi_k(T_{i,k}(x))=\phi_k(x)=(y,z)=\zeta_{i,k}(y,z)=\zeta_{i,k}(\phi_k(x)).\]

  We are left with the case of $0\leq i\leq n-k$. We now write
  \[
  T_{i,k}(x) = x_1 x_2 \dots x_{i+k} x_{i+1} x_{i+2} \dots x_n.
  \]
  Thus, if we denote $\phi_k(T_{i,k}(x))=(y,z)$, then
  \begin{align*}
    y &= x_1\dots x_k = \pref_k(x),\\
    z &= x_{k+1}-x_1,\dots,x_{k+i}-x_i,0^k,\\
    &\quad\ x_{k+i+1}-x_{i+1},\dots,x_{n}-x_{n-k}.
  \end{align*}
  This is exactly an insertion of $0^k$ after $i$ symbols in the
  second part of $\phi_k(x)$. It therefore follows that
  \[ \phi_k(T_{i,k}(x))= (y,z) = \zeta_{i,k}(\phi_k(x)),\]
  as claimed.
\end{IEEEproof}

Recalling that $\phi_k$ is a bijection between $\Z_q^n$ and $\Z_q^k\times
\Z_q^{n-k}$, together with Lemma~\ref{lem:commute} gives us the following
corollary.

\begin{corollary}
  \label{cor:domain}
  For any $x\in\Z_q^{\geq k}$, and for any sequence of non-negative integers
  $i_1,\dots,i_t$,
  \[T_{i_t,k}(\dots T_{i_1,k}(x)\dots)=
  \phi_k^{-1}(\zeta_{i_t,k}(\dots \zeta_{i_1,k}(\phi_k(x)) \dots )).\]
\end{corollary}

\begin{example}
Continuing Example~\ref{ex:diagram}, let $x=02123$, $k=t=2$, $i_1=1$,
and $i_2=0$. Then
\begin{equation*}
\begin{split}
&\phantom{{}={}} T_{0,2}(T_{1,2}(02123))\\
&=T_{0,2}(0212123)\\
&=020212123\\
&=\phi_k^{-1}((02,0010002))\\
&=\phi_k^{-1}(\zeta_{0,2}((02,10002)))\\
&=\phi_k^{-1}(\zeta_{0,2}(\zeta_{1,2}((02,102))))\\
&=\phi_k^{-1}(\zeta_{0,2}(\zeta_{1,2}(\phi_k(02123)))).
\end{split}
\end{equation*}
\end{example}

Corollary~\ref{cor:domain} paves the way to working in the
$\phi_k$-transform domain. In this domain, a tandem-duplication
operation of length $k$ translates into an insertion of a block of $k$
consecutive zeros. Conversely, a tandem-deduplication operation of
length $k$ becomes a removal of a block of $k$ consecutive zeros.

The uniqueness of the root, proved in~\cite{LeuMarMit05}, now comes
for free. In the $\phi_k$-transform domain, given
$(x,y)\in\Z_q^k\times\Z_q^*$, as long as $y$ contains a substring of
$k$ consecutive zeros, we may perform another deduplication. The
process stops at the unique outcome in which the length of every run
of zeros in $y$ is reduced modulo $k$.

This last observation motivates us to define the following operation
on a string in $\Z_q^*$. We define $\mu_k: \Z_q^*\to\Z_q^*$ which
reduces the lengths of runs of zeros modulo $k$ in the following
way. Consider a string $x\in \Z_q^*$, where
\[x=0^{m_0}w_1 0^{m_1} w_2 \dots w_t 0^{m_t},\]
where $m_i$ are non-negative integers, and $w_1,\dots,w_t \in
\Z_q\setminus\mathset{0}$, i.e., $w_1,\dots,w_t$ are single non-zero
symbols.  We then define
\[ \mu_k(x) = 0^{m_0 \bmod k}w_1 0^{m_1 \bmod k} w_2 \dots w_t 0^{m_t\bmod k}.\]
For example, for $z=0010002$,
\[ \mu_2(z) = 102.\]
Additionally, we define
\[ \sigma_k(x) = \parenv{\floorenv{\frac{m_0}{k}},\floorenv{\frac{m_1}{k}},\dots,\floorenv{\frac{m_t}{k}}} \in (\N\cup\mathset{0})^*\]
and call $\sigma(x)$ the \emph{zero signature} of $x$. For $z$ given above, 
\[ \sigma_2(z) = (1,1,0).\]
We note that $\mu_k(x)$ and $\sigma(x)$ together uniquely determine
$x$.

We also observe some simple properties. First, the Hamming weight of a
vector, denoted $\wt_H$, counts the number of non-zero elements in a
vector. By definition we have for every $x\in\Z_q^n$,
\[ \wt_H(x)=\wt_H(\mu_k(x)).\]
Additionally, the length of the vector $\sigma_k(x)$, denoted
$\abs{\sigma_k(x)}$, is given by
\begin{equation}
  \label{eq:lensig}
  \abs{\sigma_k(x)} = \wt_H(x) + 1 =\wt_H(\mu_k(x))+1.
\end{equation}
Note that for $z=0010002$ as above, we have
\[\abs{\sigma_2(z)} = 3 = \wt_H(z) + 1 =\wt_H(102)+1.\]

Thus, our
previous discussion implies the following corollary.

\begin{corollary}
  \label{cor:uniqroot}
  For any string $x\in\Z_q^{\geq k}$,
  \[ R_k(x)=\mathset{\left. \phi_k^{-1}(y,\mu_k(z)) ~\right|~ \phi_k(x)=(y,z) }.\]
\end{corollary}

We recall the definition of the $(0,k-1)$-RLL system over $\Z_q$ (for
example, see~\cite{LinMar85,Imm91}). It is defined as the set of all
finite strings over $\Z_q$ that do not contain $k$ consecutive
zeros. We denote this set as $C_{\rll_q(0,k-1)}$. In our notation,
\[ C_{\rll_q(0,k-1)} = \mathset{\left. x\in \Z_q^* ~\right|~ \sigma_k(x)\in 0^*}.\]
By convention, $C_{\rll_q(0,k-1)}\cap\Z_q^{0}=\mathset{\epsilon}$. The
following is another immediate corollary.

\begin{corollary}
  \label{cor:rll}
  For all $n\geq k$,
  \[ \irr_k(n) = \mathset{\left. \phi_k^{-1}(y,z) ~\right|~ y\in\Z_q^k, z\in C_{\rll_q(0,k-1)}\cap\Z_q^{n-k}}.\]
\end{corollary}
\begin{IEEEproof}
  The proof is immediate since $x$ is irreducible iff no deduplication action
  may be applied to it. This happens iff for $\phi_k(x)=(y,z)$, $z$ does
  not contain $k$ consecutive zeros, i.e., $z\in C_{\rll_q(0,k-1)}\cap \Z_q^{n-k}$.
\end{IEEEproof}

Given two strings, $x,x'\in \Z_q^{\geq k}$, we say $x$ and $x'$ are
$k$-congruent, denoted $x\sim_k x'$, if $R_k(x)=R_k(x')$. It is easily
seen that $\sim_k$ is an equivalence relation. 

\begin{corollary}
  \label{cor:cong}
  Let $x,x'\in\Z_q^*$ be two strings, and denote $\phi_k(x)=(y,z)$ and
  $\phi_k(x')=(y',z')$. Then $x\sim_k x'$ iff $y=y'$ and $\mu_k(z)=\mu_k(z')$.
\end{corollary}
\begin{IEEEproof}
  This is immediate when using Corollary~\ref{cor:uniqroot} to express
  the roots of $x$ and $x'$.
\end{IEEEproof}

\begin{example}
For instance, $02123$, $0212323$, $0212123$, and $020212123$ are all
2-congruent, since they have the unique root $02123$. In the
$\phi_2$-transform domain, for each sequence $x$ in the preceding
list, if we let $\phi_2(x)=(y,z)$, then $y=02$ and $\mu_2(z)=102$.
\end{example}

The following lemma appeared in~\cite[Proposition 2]{LeuMarMit05}. We
restate it and give an alternative proof.

\begin{lemma}
  \label{lem:descone}
  For all $x,x'\in\Z_q^{\geq k}$, we have
  \[ D_k^*(x)\cap D_k^*(x') \neq \emptyset\]
  if and only if $x \sim_k x'$.
\end{lemma}
\begin{IEEEproof}
  In the first direction, assume $x\not\sim_k x'$. By the uniqueness of
  the root, let us denote $R_k(x)=\mathset{u}$ and
  $R_k(x')=\mathset{u'}$, with $u\neq u'$. If there exists $w\in
  D_k^*(x)\cap D_k^*(x')$, then $w$ is a descendant of both $u$ and
  $u'$, therefore $u,u'\in R_k(w)$, which is a contradiction. Hence,
  no such $w$ exists, i.e., $D_k^*(x)\cap D_k^*(x')=\emptyset$.

  In the other direction, assume $x\sim_k x'$. We construct a word
  $w\in D_k^*(x)\cap D_k^*(x')$. Denote $\phi_k(x)=(y,z)$ and
  $\phi_k(x')=(y',z')$. By Corollary~\ref{cor:cong} we have
  \begin{align*}
    y & = y' ,\\
    \mu_k(z)&=\mu_k(z').
  \end{align*}
  Let us then denote
  \begin{align*}
    z &= 0^{m_0}v_1 0^{m_1} v_2 \dots v_t 0^{m_t}, \\
    z' &= 0^{m'_0}v_1 0^{m'_1} v_2 \dots v_t 0^{m'_t},
  \end{align*}
  with $v_i$ a non-zero symbol, and
  \[ m_i \equiv m'_i \pmod{k},\]
  for all $i$.  We now define
  \[ z'' = 0^{\max(m_0,m'_0)}v_1 0^{\max(m_1,m'_1)} v_2 \dots v_t 0^{\max(m_t,m'_t)}.\]
  Since $z''$ differs from $z$ and $z'$ by insertion of blocks of $k$
  consecutive zeros, it follows that
  \[ w = \phi_k^{-1}(y,z'') \in  D_k^*(x)\cap D_k^*(x'),\]
  which completes the proof.
\end{IEEEproof}

We now turn to constructing error-correcting codes. The first
construction is for a code capable of correcting all errors.
\begin{construction}
  \label{con:uniallerr}
  Fix $\Sigma=\Z_q$ and $k\geq 1$. For any $n\geq k$ we construct
  \[C=\bigcup_{i=0}^{\floorenv{n/k}-1} \mathset{ \left. \phi_k^{-1}(y,z0^{ki}) ~\right|~ \phi_k^{-1}(y,z)\in\irr_k(n-ik)}.\]
\end{construction}
\begin{theorem}
  \label{th:unicode}
  The code $C$ from Construction~\ref{con:uniallerr} is an
  $(n,M;*)_k$ code, with
  \[ M = \sum_{i=0}^{\floorenv{n/k}-1} q^k M_{\rll_q(0,k-1)}(n-(i+1)k).\]
  Here $M_{\rll_q(0,k-1)}(m)$ denotes the number of strings of length
  $m$ which are $(0,k-1)$-RLL over $\Z_q$, i.e.,
  \[ M_{\rll_q(0,k-1)}(m) = \abs{ C_{\rll_q(0,k-1)} \cap \Z_q^m }.\]
\end{theorem}
\begin{IEEEproof}
  The size of the code is immediate, by
  Corollary~\ref{cor:rll}. Additionally, the roots of distinct
  codewords are distinct as well, since we constructed the code from
  irreducible words with blocks of $k$ consecutive zeros appended to
  their end. Thus, by Lemma~\ref{lem:descone}, the descendant cones of
  distinct codewords are disjoint.
\end{IEEEproof}

We can say more about the size of the code we constructed.

\begin{theorem}
  The code $C$ from Construction~\ref{con:uniallerr} is optimal, i.e.,
  it has the largest cardinality of any $(n;*)_k$ code.
\end{theorem}
\begin{IEEEproof}
  By Lemma~\ref{lem:descone}, any two distinct codewords of an
  $(n;*)_k$ code must belong to different equivalence classes of
  $\sim_k$.  The code $C$ of Construction~\ref{con:uniallerr} contains
  exactly one codeword from each equivalence class of $\sim_k$, and
  thus, it is optimal.
\end{IEEEproof}

The code $C$ from Construction~\ref{con:uniallerr} also allows a
simple decoding procedure, whose correctness follows from Corollary
\ref{cor:uniqroot}. Assume a word $x'\in\Z_q^{\geq k}$ is received,
and let $\phi_k(x') = (y',z')$. The decoded word is simply
\begin{equation}\label{eq:decode} \tilde{x} = \phi_k^{-1}(y',\mu_k(z') 0^{n-k-\abs{\mu_k(z')}}),\end{equation}
where $n$ is the length of the code $C$. In other words, the decoding
procedure recovers the unique root of the received $x'$, and in the
$\phi_k$-transform domain, pads it with enough zeros.

\begin{example}
Let $n=4$, $q=2$, and $k=1$. By inspection, the code $C$ of
Construction~\ref{con:uniallerr} can be shown to equal
\[C=\mathset{\underline{0}000,\underline{01}11,\underline{010}0,\underline{0101},\underline{1}111,\underline{10}00,\underline{101}1,\underline{1010}},\]
where in each codeword the k-irreducible part is underlined. As an
example of decoding, both $01100$ and $01000$ decode to
$0100$. Specifically for the former case, $x'=01100$, we have
$\phi_k(x')=(y',z')=(0,1010)$. So $\mu_k(z')=11$ and
\[\tilde x =\phi_k^{-1}(0,110)=0100.\]
\end{example}

Encoding may be done in any of the many various ways for encoding
RLL-constrained systems. The reader is referred to
\cite{LinMar85,Imm91} for further reading. After encoding the
RLL-constrained string $z$, a string $y\in\Z_q^k$ is added, and
$\phi_k^{-1}$ employed, to obtain a codeword.

Finally, the asymptotic rate of the code family may also be obtained,
thus, obtaining the capacity of the channel.

\begin{corollary}
  For all $q\geq 2$ and $k\geq 1$,
  \[ \ccap_q(*)_k = \ccap(\rll_q(0,k-1)),\]
where $\ccap(\rll_q(0,k-1))$ is the capacity of the $q$-ary
$(0,k-1)$-RLL constrained system.
\end{corollary}
\begin{IEEEproof}
We use $C_n$ to denote the code from Construction~\ref{con:uniallerr},
where the subscript $n$ is used to denote the length of the code. It
is easy to see that for $n\geq k$,
\[ q^k M_{\rll_q(0,k-1)}(n-k) \leq \abs{C_n} \leq nq^k M_{\rll_q(0,k-1)}(n-k).\]
Then by standard techniques~\cite{LinMar85} for constrained coding,
\begin{align*}
  \lim_{n\to\infty}\frac{1}{n}\log_2\abs{C_n} &= \ccap(\rll_q(0,k-1)) \\
  &= \log_2 \lambda(A_q(k-1)),
\end{align*}
where $\lambda(A_q(k-1))$ is the largest eigenvalue of the $k\times k$ matrix $A_q(k-1)$ defined as
\begin{equation}
  \label{eq:mat0k}
  A_q(k-1) =
  \begin{pmatrix}
    q-1    & 1 &   &        & \\
    q-1    &   & 1 &        & \\
    \vdots &   &   & \ddots & \\
    q-1    &   &   &        & 1\\
    q-1    &   &   &        &
  \end{pmatrix}.
\end{equation}
\end{IEEEproof}

As a side note, we comment that an asymptotic (in $k$) expression for
the capacity may be given by
\begin{equation}
  \label{eq:cap0k}
  \ccap(\rll_q(0,k)) = \log_2 q - \frac{(q-1)\log_2 e}{q^{k+2}}(1+o(1)).
\end{equation}
This expression agrees with the expression for the binary case $q=2$
mentioned in~\cite{KatZeg99} without proof or reference. For
completeness, we bring a short proof of this claim in the appendix.

Having considered $(n,M;*)_k$ codes, we now turn to study $(n,M;t)_k$
codes for $t\in\N\cup\mathset{0}$. We note that $\Z_q^n$ is an optimal
$(n,q^n;0)_k$ code. Additionally, any $(n,M;*)_k$ code is trivially
also an $(n,M;t)_k$ code, though not necessarily optimal.

We know by Lemma~\ref{lem:descone} that the descendant cones of two
words overlap if and only if they are $k$-congruent. Thus, the
strategy for constructing $(n,M;*)_k$ codes was to pick single
representatives of the equivalence classes of $\sim_k$ as
codewords. However, the overlap that is guaranteed by Lemma
\ref{lem:descone} may require a large amount of duplication
operations. If we are interested in a small enough value of $t$, then
an $(n,M;t)_k$ code may contain several codewords from the same
equivalence class. This observation will be formalized in the
following, by introducing a metric on $k$-congruent words, and
applying this metric to pick $k$-congruent codewords.

Fix a length $n\geq 1$, and let $x,x'\in\Z_q^n$, $x\sim_k x'$, be two
$k$-congruent words of length $n$. We define the distance between
$x$ and $x'$ as
\[ d_k(x,x') = \min\mathset{ t\geq 0 ~\left|~ D_k^t(x) \cap D_k^t(x')\neq \emptyset \right. }.\]
Since $x$ and $x'$ are $k$-congruent, Lemma~\ref{lem:descone} ensures
that $d_k$ is well defined.

\begin{lemma}
\label{lem:metric}
  Let $x,x'\in\Z_q^n$, $x\sim_k x'$, be two $k$-congruent strings. Denote
  $\phi_k(x)=(y,z)$ and $\phi_k(x')=(y,z')$. Additionally,
  let
  \begin{align*}
    \sigma_k(z) = (s_0,s_1,\dots,s_r),\\
    \sigma_k(z') = (s'_0,s'_1,\dots,s'_r).
  \end{align*}
  Then
  \[ d_k(x,x') = \sum_{i=0}^{r} \abs{s_i-s'_i} =
  d_{\ell_1}(\sigma_k(z),\sigma_k(z')),\] where $d_{\ell_1}$ stands
  for the $\ell_1$-distance function.
\end{lemma}
\begin{IEEEproof}
  Let $x$ and $x'$ be two strings as required. By Corollary~\ref{cor:cong} we indeed have $y=y'$, and $\mu_k(z)=\mu_k(z')$. In
  particular, the length of the vectors of the zero signatures of $z$
  and $z'$ are the same,
  \[ \abs{\sigma_k(z)}=\abs{\sigma_k(z')}=r+1. \]
  We now observe that the action of a $k$-tandem duplication on $x$
  corresponds to the addition of a standard unit vector $e_i$ (an
  all-zero vector except for the $i$th coordinate which equals $1$) to
  $\sigma_k(z)$.

  Let $\tilde{x}$ denote a vector that is a descendant both of $x$ and
  $x'$, and that requires the least number of $k$-tandem duplications
  to reach from $x$ and $x'$. If we denote
  $\phi_k(\tilde{x})=(\tilde{y},\tilde{z})$, then we have
  \begin{align*}
    \tilde{y} &= y = y',\\
    \mu_k(\tilde{z})&=\mu_k(z)=\mu_k(z'),\\
    \sigma_k(\tilde{z}) &= (\max(s_0,s'_0),\dots,\max(s_r,s'_r)).
  \end{align*}
  Thus,
  \begin{align*}
    d_k(x,x') &= \sum_{i=0}^r (\max(s_i,s'_i)-s_i) \\
    &= \sum_{i=0}^r (\max(s_i,s'_i)-s'_i)\\
    &= \sum_{i=0}^{r} \abs{s_i-s'_i} =
    d_{\ell_1}(\sigma_k(z),\sigma_k(z')).
  \end{align*}
\end{IEEEproof}

From Lemma~\ref{lem:metric} we also deduce that $d_k$ is a metric over
any set of $k$-congruent words of length $n$.

The following theorem shows that a code is $(n;t)_k$ if and only if
the zero signatures of the $z$-part of $k$-congruent codewords in the
$\phi_k$-transform domain, form a constant-weight code in the
$\ell_1$-metric with distance at least $t+1$. We recall that the
$\ell_1$-metric weight of a vector $s=s_1 s_2\dots s_n\in\Z^n$ is
defined as the $\ell_1$-distance to the zero vector, i.e.,
\[\wt_{\ell_1}(s)=\sum_{i=1}^n \abs{s_i}.\]

\begin{theorem}
  \label{th:l1code}
  Let $C\subseteq \Z_q^n$, $n\geq k$, be a subset of size $M$. Then
  $C$ is an $(n,M;t)_k$ code if and only if for each $y\in\Z_q^k$,
  $z\in\Z_q^{n-k}$, the following sets
  \begin{align*}
    C(y,z) =\Big\{ \sigma_k(z') ~\Big|~ &z'\in\Z_q^{n-k}, \mu_k(z)=\mu_k(z'),  \\
      &\phi_k^{-1}(y,z')\in C \Big\}
  \end{align*}
  are constant-weight $(n(y,z),M(y,z),t+1)$ codes in the
  $\ell_1$-metric, with constant weight
  \[\wt_{\ell_1}(\sigma(z))=\frac{n-k-\abs{\mu_k(z)}}{k},\]
  and length
  \[n(y,z) = \wt_H(z)+1=\wt_H(\mu_k(z))+1,\]
  where $\wt_H$ denotes the Hamming weight.
\end{theorem}
\begin{IEEEproof}
  In the first direction, let $C$ be an $(n,M;t)_k$ code. Fix $y$ and
  $z$, and consider the set $C(y,z)$. Assume to the contrary that
  there exist distinct $\sigma_k(z'),\sigma_k(z'')\in C(y,z)$,
  $z',z''\in\Z_q^{n-k}$, such that
  $d_{\ell_1}(\sigma_k(z'),\sigma_k(z'')) \leq t$.

  The length of the code, $n(y,z)$, is obvious given
  \eqref{eq:lensig}. We note that $\sigma_k(z')\neq \sigma_k(z'')$
  implies $z'\neq z''$.  By definition, we have
  \[ \mu_k(z)=\mu_k(z')=\mu_k(z'').\]
  Thus,
  \begin{align*}
    \wt_{\ell_1}(\sigma(z)) & = \wt_{\ell_1}(\sigma(z')) = \wt_{\ell_1}(\sigma(z''))\\
    &= \frac{n-k-\abs{\mu_k(z)}}{k},
  \end{align*}
  where $\abs{\mu_k(z)}$ denotes the length of the vector $\mu_k(z)$.
  Additionally, the two codewords
  \[ c'=\phi_k^{-1}(y,z')\in C \qquad\text{and}\qquad
  c''=\phi_k^{-1}(y,z'')\in C\]
  are $k$-congruent and distinct. By Lemma~\ref{lem:metric},
  \begin{equation}
    \label{eq:desint}
    d_k(c',c'') = d_{\ell_1}(\sigma_k(z'),\sigma_k(z''))\leq t.
  \end{equation}
  However, that contradicts the code parameters since we have
  \eqref{eq:desint} imply $D_k^t(c')\cap D_k^t(c'')\neq \emptyset$,
  whereas in an $(n,M;t)_k$ code, the $t$-descendant cones of distinct
  codewords have an empty intersection.

  In the other direction, assume that for every choice of $y$ and $z$,
  the corresponding $C(y,z)$ is a constant-weight code with minimum
  $\ell_1$-distance of $t+1$. Assume to the contrary $C$ is not an
  $(n,M;t)_k$ code. Therefore, there exist two distinct codewords,
  $c',c''\in C$ such that $d_k(c',c'')\leq t$.

  By Lemma~\ref{lem:descone} we conclude that $c'$ and $c''$ are
  $k$-congruent. Thus, there exist $y\in\Z_q^k$ and $z\in\Z_q^{n-k}$
  ($z$ is not necessarily unique) such that,
  \begin{align*}
    \phi_k(c')&=(y,z') \\
    \phi_k(c'')&=(y,z'')\\
    \mu_k(z)&=\mu_k(z')=\mu_k(z'').
  \end{align*}
  We can now use Lemma~\ref{lem:metric} and obtain
  \[ d_{\ell_1}(\sigma_k(z'),\sigma_k(z''))=d_k(c',c'') \leq t,\]
  which contradicts the minimal distance of $C(y,z)$.
\end{IEEEproof}

With the insight given by Theorem~\ref{th:l1code} we now give a
construction for $(n,M;t)_k$ codes.

\begin{construction}
  \label{con:uniterr}
  Fix $\Sigma=\Z_q$, $k\geq 1$, $n\geq k$, and $t\geq 0$. Furthermore, for all
  \begin{align*}
    1 & \leq m\leq n-k+1,\\
    0 & \leq w\leq \floorenv{\frac{n-k}{k}},
  \end{align*}
  fix $\ell_1$-metric codes over $\Z_q$, denoted $C_1(m,w)$, which are
  of length $m$, constant $\ell_1$-weight $w$, and minimum
  $\ell_1$-distance $t+1$. We construct
  \[
    C = \bigg\{ \phi_k^{-1}(y,z) ~\bigg|~ y\in\Z_q^k, z\in\Z_q^{n-k},
      \sigma_k(z)\in C_1\parenv{\wt_H(\mu_k(z))+1,\frac{n-k-\abs{\mu_k(z)}}{k}}\bigg\}.
  \]
\end{construction}

\begin{corollary}
  The code $C$ from Construction~\ref{con:uniterr} is an $(n,M;t)_k$ code.
\end{corollary}
\begin{IEEEproof}
  Let $c,c'\in C$ be two $k$-congruent codewords, i.e.,
  $\phi_k(c)=(y,z)$, $\phi_k(c')=(y,z')$, and $\mu_k(z)=\mu_k(z')$. It
  follows, by construction, that $\sigma_k(z)$ and $\sigma_k(z')$
  belong to the same $\ell_1$-metric code with minimum
  $\ell_1$-distance at least $t+1$. By Theorem~\ref{th:l1code}, $C$ is
  an $(n,M;t)_k$ code.
\end{IEEEproof}

Due to Theorem~\ref{th:l1code}, a choice of optimal $\ell_1$-metric
codes in Construction~\ref{con:uniterr} will result in optimal
$(n,M;t)_k$ codes. We are unfortunately unaware of explicit
construction for such codes. However, we may deduce such a
construction from codes for the similar Lee metric (e.g.,
\cite{RotSie94}), while applying a standard averaging argument for
inferring the existence of a constant-weight code. We leave the
construction of such codes for a future work.


\section{$\leq k$-Tandem-Duplication Codes}\label{sec:le-k-tandem}

In this Section, we consider error-correcting codes that correct
duplications of length at most $k$, which correspond to $\stan_{\leq
  k}$. In particular, we present constructions for codes that can
correct any number of duplications of length $\le 3$ as well as a
lower bound on the capacity of the corresponding channel. In the case
of duplications of length $\leq 2$ we give optimal codes, and obtain
the exact capacity of the channel.

It is worth noting that the systems $\stan_{\leq k}$ were studied in
the context of formal languages~\cite{LeuMarMit05} and also in the
context of coding and information
theory~\cite{JaiFarBru15}. In~\cite{LeuMarMit05}, it was shown that
$\stan_{\leq k}$, with $k\geq 4$, is not a regular language for
alphabet size $\abs{\Sigma} \geq 3$. However, it was proved
in~\cite{JaiFarBru15} that $\stan_{\leq 3}$ is indeed a regular
language irrespective of the starting string and the alphabet size.

In this paper, we will show that strings that can be generated by
bounded tandem string-duplication systems with maximum duplication
length $3$ have a unique duplication root, a fact that will be useful
for our code construction. Theorem~\ref{th:unqroot3} formalizes this
statement. We begin with the following definition.

\begin{definition}
Let two squares $y_1 = \alpha\alpha\in \Sigma^+$ and $y_2 = \beta\beta
\in \Sigma^+$ appear as substrings of some string $u \in \Sigma^*$, i.e.,
\[ u=x_1 y_1 z_1 = x_2 y_2 z_2,\]
with $\abs{x_1}=i$, $\abs{x_2}=j$.  We say $y_1$ and $y_2$ are
\emph{overlapping squares in $u$} if the following conditions both
hold:
\begin{enumerate}
\item $i \leq j \leq i+2\abs{\alpha}-1$ or $j \leq i \leq j+2\abs{\beta}-1$.
\item If $i =j$, then $\alpha \neq \beta$.
\end{enumerate}
\end{definition}
\begin{example}
\newlength\lunderset
\newlength\rulethick
\lunderset=1.5pt\relax
\rulethick=.8pt\relax
\def\stackalignment{l}
\newcommand\nunderline[3][1]{\setbox0=\hbox{#2}%
  \stackunder[#1\lunderset-\rulethick]{\strut#2}{\color{#3}\rule{\wd0}{\rulethick}}}
Consider the sequence $u$,
\[u=0\,1\lefteqn{\overbrace{\phantom{\,2\,3\,2\,3\,4\,5\,2\,4\,5\,2\,3\,2\,3\,4\,5\,2\,4\,5\,}}^{\alpha\alpha}}\underbrace{2\,3\,2\,3}_{\beta_1\beta_1}\underbrace{4\,5\,2\,4\,5\,2}_{\beta_2\beta_2}3\,2\,3\,4\,5\underbrace{2\,4\,5\,6\,2\,4\,5\,6}_{\beta_3\beta_3}7,\]
where $\alpha\alpha$ and $\beta_i\beta_i$ for each $i\in\mathset{1,2,3}$ are overlapping squares.
\end{example}

The following theorem shows that every word has a unique root under
tandem deduplication of length up to $3$.

\begin{theorem}
  \label{th:unqroot3}
  For any $z\in \Sigma^*$ we have $\abs{R_{\leq 3}(z)}=1$.
\end{theorem}

\begin{IEEEproof}
  Fix some $z\in\Sigma^*$, and assume $z$ has exactly $m$ distinct roots,
  $R_{\leq 3}(z)=\mathset{y_1,y_2,\dots,y_m}$. Let us assume to the
  contrary that $m \geq 2$.

  Let us follow a deduplication sequence starting at $x_0=z$. At each
  step, we deduplicate $x_i \dedup_{\leq 3} x_{i+1}$, and we must have
  $\abs{R_{\leq 3}(x_i)}\geq \abs{R_{\leq 3}(x_{i+1})}$. At each step,
  out of the possible immediate ancestors of $x_i$, we choose
  $x_{i+1}$ to be one with $\abs{R_{\leq 3}(x_{i+1})}\geq 2$ if
  possible. Since the end-point of a deduplication process is an
  irreducible sequence, we must reach a sequence $x$ in the
  deduplication process with the following properties:
  \begin{enumerate}
  \item
    $z \dedup_{\leq 3}^* x$
  \item
    $\abs{R_{\leq 3}(x)} \geq 2$
  \item
    For each $x'\in\Sigma^*$ such that $x \dedup_{\leq 3} x'$, 
    $\abs{R_{\leq 3}(x')}=1$.
  \item
    There exist $v,w\in\Sigma^*$ such that $x\dedup_{\leq 3} v$ and
    $x\dedup_{\leq 3} w$ with $\abs{R_{\leq 3}(v)}=\abs{R_{\leq 3}(w)}=1$.
  \item
    $R_{\leq 3}(v)=\mathset{y_i}\neq \mathset{y_j}=R_{\leq 3}(w)$.
  \end{enumerate}
  Intuitively, in the deduplication process starting from $z$, we
  reach a sequence $x$ with more than one root, but any following
  single deduplication moves us into a single descendant cone of one
  of the roots of $z$. We note that all ancestors of $v$ must have a
  single root $y_i$, and all ancestors of $w$ must have a single root
  $y_j$.

  Thus, $x$ must contain a square $u_v u_v$ whose deduplication
  results in $v$, and a square $u_w u_w$ whose deduplication results
  in $w$.  We contend that the squares $u_v u_v$ and $u_w u_w$ overlap.
  Otherwise, if $u_v u_v$ and $u_w u_w$ do not overlap in $x$, we may
  deduplicate them in any order to obtain the same result. Hence,
  there exists $t \in \Sigma^*$ such that $v\dedup_{\leq 3} t$ and
  $w\dedup_{\leq 3} t$. But then, since $t$ is an ancestor both of $v$
  and $w$,
  \[ \mathset{y_i}=R_{\leq 3}(v) = R_{\leq 3}(t) = R_{\leq 3}(w) = \mathset{y_j},\]
  a contradiction.


  We now know that $u_v u_v$ and $u_w u_w$ must overlap. We also note
  $\abs{u_v},\abs{u_w}\leq 3$.  Let $a,b,c\in\Sigma$ be three distinct
  symbols. If the alphabet is smaller, then some of the cases below
  may be ignored, and the proof remains the same. We use brute force
  to enumerate the following cases: (each string describes the
  shortest subsequence that contains the overlapping squares)
\begin{enumerate}
\item $\abs{u_v} = 1, \abs{u_w} = 1:$ $aaa$.
\item $\abs{u_v} = 1, \abs{u_w} = 2:$ $aaaaa$, $aabab$.
\item $\abs{u_v} = 1, \abs{u_w} = 3:$ $aaaaaa$, $aaaaaaa$, $aabaaba$, $abaaba$, $aabcabc$.
\item $\abs{u_v} = 2, \abs{u_w} = 2:$ $aaaaa$, $ababab$, $ababbbb$, $ababa$, $bcbcaca$.
\item $\abs{u_v} = 2, \abs{u_w} = 3:$ $aaaaaa$, $aaaaaaa$, $aaaaaaaa$, $aaaaaaaaa$, $abaabaaaa$, $abaababa$, $abaabab$, $abcabcccc$, $abcabcaca$, $abcabcbcb$, $abcabcbc$.
\item $\abs{u_v} = 3, \abs{u_w} = 3:$ $aaaaaaa$, $aaaaaaaa$, $aaaaaaaaa$, $aaaaaaaaaa$, $aaaaaaaaaaa$, $abaabaaaaaa$, $abaababaaba$, $abaabacaaca$, $abaababcabc$, $abaabacbacb$, $abaabaabaa$, $abaababbab$, $abaabacbac$, $abaabaaba$, $abaabaab$, $abaabaa$, $abcabcaacaa$, $abcabcbbcbb$, $abcabcbccbc$, $abcabcaccac$, $abcabccbccb$, $abcabccacca$, $abcabccbcc$, $abcabcbbcb$, $abcabcabca$, $abcabcabc$, $abcabcab$, $abcabca$.
\end{enumerate}
All other cases left are symmetric (by relabeling the alphabet
symbols) to one of the above listed case.  For example, if $u_v = abc$
and $u_w = cbc$, the corresponding string appears in case 6) as
$abcabcbccbc$.  It is tedious, yet easy, to check that each of the
above listed cases has a unique root if deduplication of maximum
length $3$ is allowed. In the above example, indeed, the only possible
root is $abc$,
\begin{align*}
\underline{abcabc}bccbc &\dedup_{\leq 3} abcbccbc \dedup_{\leq 3}^* abc,\\
abcab\underline{cbccbc} &\dedup_{\leq 3} abcabcbc \dedup_{\leq 3}^* abc.
\end{align*}

Let $x = \alpha \beta \gamma \in\Sigma^*$, where $\beta$ covers
exactly the overlapping squares, and is one of the above listed
cases. Then, by deduplication of $u_v u_v$ from $\beta$ in $x$, we get
$v$, and by deduplication of $u_w u_w$ from $\beta$ in $x$, we get
$w$. However, since $\beta$ has a unique root, we may deduplicate $v$
and $w$ to the same word $t=\alpha \beta' \gamma \in\Sigma^*$, where
$R(\beta)=\mathset{\beta'}$, i.e., $\beta'$ is the unique root of
$\beta$. Thus, $t$ is an ancestor of both $v$ and $w$.  Again,
\[ \mathset{y_i}=R_{\leq 3}(v) = R_{\leq 3}(t) = R_{\leq 3}(w) = \mathset{y_j},\]
which is a contradiction.
\end{IEEEproof}
\begin{corollary}
  \label{cor:unqroot2}
  For any $z\in \Sigma^*$ we also have $\abs{R_{\leq k}(z)}=1$ for
  $k=1,2$.
\end{corollary}

In a similar fashion to the previous section, we define the following
relation. We say $x,x'\in\Sigma^*$ are $\leq 3$-congruent, denoted $x
\sim_{\leq 3} x'$, if $R_{\leq 3}(x)=R_{\leq 3}(x')$. Clearly
$\sim_{\leq 3}$ is an equivalence relation. Having shown any sequence
has a unique root when duplicating up to length $3$, we obtain the
following corollary.

\begin{corollary}
  \label{cor:simupto3}
  For any two words $x,x'\in\Sigma^*$, if
  \[D_{\leq 3}^*(x)\cap D_{\leq 3}^*(x') \neq \emptyset\]
  then $x\sim_{\leq 3} x'$.
\end{corollary}

We note that unlike Lemma~\ref{lem:descone}, we do not have
$x\sim_{\leq 3} x'$ necessarily imply that their descendant cones
intersect. Here is a simple example illustrating this case. Fix $q=3$,
and let $x=012012$ and $x'=001122$. We note that $x\sim_{\leq 3} x'$,
since
\[ R_{\leq 3}(x)=R_{\leq 3}(x')=\mathset{012}.\]
However, $D_{\leq 3}^*(x)\cap D_{\leq 3}^*(x') = \emptyset$ since all
the descendants of $x$ have a $0$ to the right of a $2$, whereas all
the descendants of $x'$ do not.

We are missing a simple operator which is required to define an
error-correcting code. For any sequence $x\in\Sigma^+$, we define
its $k$-suffix-extension to be
\[ \xi_k(x) = x (\suff_1(x))^k,\]
i.e., the sequence $x$ with its last symbol repeated an extra $k$
times.

\begin{construction}
  \label{con:upto3}
  Let $\Sigma$ be some finite alphabet. The constructed code is
  \[ C = \bigcup_{i=1}^n \mathset{ \xi_{n-i}(x) ~|~ x\in\irr_{\leq 3}(i) }.\]
\end{construction}

\begin{theorem}
  \label{th:codeupto3}
  The code $C$ from Construction~\ref{con:upto3} is an ${(n,
    M;*)}_{\leq 3}$ code, where
  \[ M = \sum_{i=1}^n \abs{\irr_{\leq 3}(i)}.\]
\end{theorem}
\begin{IEEEproof}
  The parameters of the code are obvious. Since the last letter
  duplication induced by the suffix extension may be deduplicated, we
  clearly have exactly one codeword from each equivalence class of
  $\sim_{\leq 3}$. By Corollary~\ref{cor:simupto3}, the descendant
  cones of the codewords do not intersect and the code can indeed
  correct all errors.
\end{IEEEproof}

For the remainder of the section we denote by $\irr_{q;\leq 3}$ the
set of irreducible words with respect to $\dedup_{\leq 3}$ over
$\Z_q$, in order to make explicit the dependence on the size of the
alphabet. We also assume $q\geq 3$, since $q=2$ is a trivial case with
\begin{equation}
  \label{eq:irrbinary}
  \irr_{2;\leq
    3}=\mathset{0,1,01,10,010,101}.
\end{equation}

\begin{figure*}[ht]
  \begin{center}
    \psfrag{v01}{\scriptsize $10201$}
    \psfrag{v02}{\scriptsize $02010$}
    \psfrag{v03}{\scriptsize $20102$}
    \psfrag{v04}{\scriptsize $01020$}
    \psfrag{v05}{\scriptsize $01021$}
    \psfrag{v06}{\scriptsize $10210$}
    \psfrag{v07}{\scriptsize $02101$}
    \psfrag{v08}{\scriptsize $21012$}
    \psfrag{v09}{\scriptsize $10121$}
    \psfrag{v10}{\scriptsize $01210$}
    \psfrag{v11}{\scriptsize $12101$}
    \psfrag{v12}{\scriptsize $12102$}
    \psfrag{v13}{\scriptsize $21021$}
    \psfrag{v14}{\scriptsize $10212$}
    \psfrag{v15}{\scriptsize $02120$}
    \psfrag{v16}{\scriptsize $21202$}
    \psfrag{v17}{\scriptsize $12021$}
    \psfrag{v18}{\scriptsize $20212$}
    \psfrag{v19}{\scriptsize $20210$}
    \psfrag{v20}{\scriptsize $02102$}
    \psfrag{v21}{\scriptsize $21020$}
    \psfrag{v22}{\scriptsize $02012$}
    \psfrag{v23}{\scriptsize $20121$}
    \psfrag{v24}{\scriptsize $10120$}
    \psfrag{v25}{\scriptsize $01202$}
    \psfrag{v26}{\scriptsize $21201$}
    \psfrag{v27}{\scriptsize $12010$}
    \psfrag{v28}{\scriptsize $20120$}
    \psfrag{v29}{\scriptsize $01201$}
    \psfrag{v30}{\scriptsize $12012$}
    \includegraphics[scale=0.6]{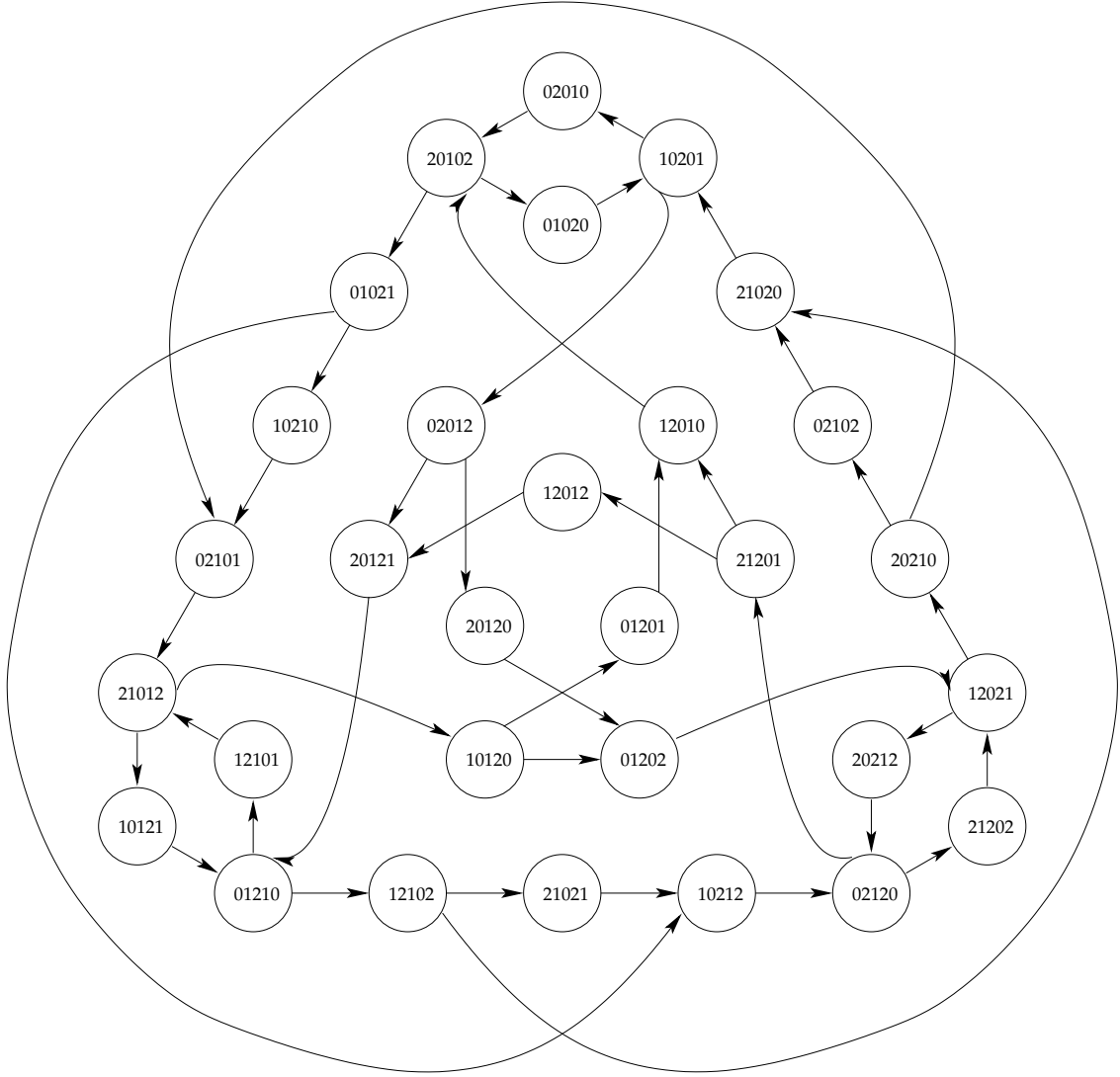}
    \end{center}
\caption{The graph $\cG'_3$ producing the set of ternary irreducible words $\irr_{3;\leq 3}$. Vertices without edges were removed as well.}
\label{fig:irr3}
\end{figure*}

We observe that $\irr_{q;\leq 3}$ is a regular language. Indeed, it is
defined by a finite set of subsequences we would like to avoid. This
set is exactly
\[\cF_q=\mathset{\left. uu\in \Z_q^* ~\right|~ 1\leq \abs{u}\leq 3}.\]
We can easily construct a finite directed graph with labeled edges
such that paths in the graph generate exactly $\irr_{q;\leq 3}$. This
graph is obtained by taking the De Bruijn graph $\cG_q=(\cV_q,\cE_q)$
of order $5$ over $\Z_q$, i.e., $\cV_q=\Z_q^5$, and edges of the form
$(a_1,a_2,a_3,a_4,a_5)\to (a_2,a_3,a_4,a_5,a_6)$, for all
$a_i\in\Z_q$. Thus, each edge is labeled with a word
$w=(a_1,a_2,a_3,a_4,a_5,a_6)\in\Z_q^6$. We then remove all edges
labeled by words $\alpha\beta\gamma\in\Z_q^6$ such that
$\beta\in\cF_q$. We call the resulting graph $\cG'_q$. It is easy
verify that each path in $\cG'_q$ generates a sequence of sliding
windows of length $6$. Reducing each window to its first letter we get
exactly $\irr_{q;\leq 3}$. An example showing $\cG'_3$ is given in
Figure~\ref{fig:irr3}. Finally, it follows that using known techniques
\cite{LinMar85}, we can calculate $\ccap(\irr_{q;\leq 3})$.

\begin{corollary}
  For all $q\geq 3$,
  \[\ccap_q(*)_{\leq 3}\geq \ccap(\irr_{q;\leq 3}).\]
\end{corollary}
\begin{IEEEproof}
  Let $M_n$ denote the size of the length $n$ code over $\Z_q$ from
  Construction~\ref{con:upto3}. By definition, $A_q(n;*)_{\leq 3}\geq
  M_n$. We note that trivially
  \[ M_n = \sum_{i=1}^{n}\abs{\irr_{q;\leq 3}(i)} \geq \abs{\irr_{q;\leq 3}(n)}.\]
  Plugging this into the definition of the capacity gives us
  the desired claim.
\end{IEEEproof}

\begin{example}
  Using the constrained system presented in Figure~\ref{fig:irr3} that
  generates $\irr_{3;\leq 3}$, we can calculate
  \[\ccap_3(*)_{\leq 3} \geq 0.347934.\]
\end{example}

Stronger statements may be given when the duplication length is upper
bounded by $2$ instead of $3$.

\begin{lemma}\label{lem:bnd_2}
For all $x, x' \in \Sigma^*$, we have
\[D_{\leq 2}^*(x)\cap D_{\leq 2}^*(x') \neq \emptyset\]
if and only if $x \sim_{\leq 2} x'.$
\end{lemma}
\begin{IEEEproof}
In the first direction, assume $x \nsim_{\leq 2} x'.$ By the
uniqueness of the root from Corollary \ref{cor:unqroot2}, let us
denote $R_{\leq 2}(x) = \mathset{u}$ and $R_{\leq 2}(x') = \mathset{u'}$, with $u
\neq u'. $ If there exists $w \in D_{\leq 2}^*(x) \cap D_{\leq
  2}^*(x')$, then $w$ is a descendant of both $u$ and $u'$, therefore
$u$ and $u' \in R_{\leq2}(w)$, which is a contradiction. Hence, no
such $w$ exists, i.e., $D_{\leq 2}^*(x) \cap D_{\leq 2}^*(x') =
\emptyset$.

In the other direction, assume $x\sim_{\leq 2} x'$. We construct a
word $w \in D_{\leq 2}^*(x) \cap D_{\leq 2}^*(x').$ Let $R_{\leq 2}(x)
= R_{\leq 2}(x') = \mathset{v}$, and denote $v=a_1a_2\dots a_m$,
where $a_i \in \Sigma$.  Consider a tandem-duplication string system
$S_{\leq 2} = (\Sigma, v, \mathcal{T}_{\leq 2}).$
Using~\cite{JaiFarBru15}, the regular expression for the language
generated by $S_{\leq 2}$ is given by
\[
a_1^+a_2^+{(a_1^+a_2^+)}^*a_3^+{(a_2^+a_3^+)}^*\dots
a_m^+{(a_{m-1}^+a_m^+)}^*.
\]
Since $x,x' \in S$, we have
\begin{align*}
x &= \prod_{i=1}^{\alpha_1}
(a_1^{p_{1i}}a_2^{q_{1i}})a_3^{q_{21}}\prod_{i=2}^{\alpha_2}
(a_2^{p_{2i}}a_3^{q_{2i}}) \\
&\quad\ \dots
a_m^{q_{(m-1)1}}\prod_{i=2}^{\alpha_{m-1}}
(a_{m-1}^{p_{(m-1)i}}a_m^{q_{(m-1)i}}),
\end{align*}
and
\begin{align*}
x' &=
\prod_{i=1}^{\beta_1}
(a_1^{e_{1i}}a_2^{f_{1i}})a_3^{f_{21}}\prod_{i=2}^{\beta_2}
(a_2^{e_{2i}}a_3^{f_{2i}})\\
&\quad\ \dots
a_m^{f_{(m-1)1}}\prod_{i=2}^{\beta_{m-1}}
(a_{m-1}^{e_{(m-1)i}}a_m^{f_{(m-1)i}}),
\end{align*}
where $\prod$ represents concatenation and $p_{ji}, q_{ji},
e_{ji}, f_{ji}, \alpha_j,\beta_j \geq 1$.  Now, it is easy to observe
that we can obtain
\[w = \prod_{i=1}^{\gamma_1}
(a_1^{g_1}a_2^{h_1})a_3^{h_2}\prod_{i=2}^{\gamma_2}
(a_2^{g_2}a_3^{h_2})\dots a_m^{h_{m-1}}\prod_{i=2}^{\gamma_{m-1}}
(a_{m-1}^{g_{m-1}}a_m^{h_{m-1}})\]
by doing tandem duplication of
length up to $2$ on $x$ and $x'$, and choosing $\gamma_j = \max\mathset{\alpha_j,\beta_j}$, $g_j = \max_{i}\mathset{p_{ji}, e_{ji}}$, and $h_j = \max_i\mathset{q_{ji},f_{ji}}$. Note, $p_{ji}$ and $q_{ji}$ are assumed to be $0$ for $i >
\alpha_j$ and $e_{ji}$ and $f_{ji}$ are assumed to be $0$ for $i >
\beta_{j}$. Thus, $w \in D_{\leq 2}(x) \cap D_{\leq 2}(x')$.
\end{IEEEproof}
\begin{construction}
  \label{con:upto2}
  Let $\Sigma$ be some finite alphabet. The constructed code is
  \[ C = \bigcup_{i=1}^n \mathset{ \xi_{n-i}(x) ~|~ x\in\irr_{\leq 2}(i) }.\]
\end{construction}
\begin{theorem}
  \label{th:codeupto2}
  The code $C$ from Construction~\ref{con:upto2} is an optimal ${(n,
    M;*)}_{\leq 2}$ code, where
  \[ M = \sum_{i=1}^n \abs{\irr_{\leq 2}(i)}.\]
\end{theorem}
\begin{IEEEproof}
  The correctness of the parameters follows the same reasoning as the
  proof of Theorem \ref{th:codeupto3}. By Lemma~\ref{lem:bnd_2}, any two
  distinct codewords of an $(n;*)_{\leq 2}$ code must belong to
  different equivalence classes of $\sim_{\leq 2}$.  The code $C$ of
  Construction~\ref{con:upto2} contains exactly one codeword from each
  equivalence class of $\sim_{\leq 2}$, and thus, it is optimal.
\end{IEEEproof}

\begin{corollary}
  For all $q\geq 3$,
  \[\ccap_q(*)_{\leq 2}= \ccap(\irr_{q;\leq 2}).\]
\end{corollary}
\begin{IEEEproof}
  Let $M_n$ denote the size of the length $n$ code over $\Z_q$ from
  Construction~\ref{con:upto2}. By definition, $A_q(n;*)_{\leq 2}\geq
  M_n$. We note that trivially
  \[ M_n = \sum_{i=1}^{n}\abs{\irr_{q;\leq 2}(i)} \geq \abs{\irr_{q;\leq 2}(n)}.\]
  Additionally, $\abs{\irr_{q;\leq 2}}(n)$ is monotone increasing in $n$
  since any irreducible length-$n$ word $x$ may be extended to an irreducible
  word of length $n+1$ by adding a letter that is not one of the last two
  letters appearing in $x$. Thus,
  \[ M_n = \sum_{i=1}^{n}\abs{\irr_{q;\leq 2}(i)} \leq n\abs{\irr_{q;\leq 2}(n)}.\]
  Plugging this into the definition of the capacity gives us
  the desired claim.
\end{IEEEproof}

\section{Duplication Roots}\label{sec:roots}

In Section \ref{sec:k-tandem}, we stated that if the duplication
length is uniform (i.e., a constant $k$ ), then every sequence has a
unique root. Further in Section \ref{sec:le-k-tandem}, we proved in
Theorem \ref{th:unqroot3} that if the duplication length is bounded by
$3$ (i.e. $\leq 3$), then again every sequence will have a unique
root. In fact, the two cases proved in the paper are the only cases of
tandem-duplication channels that have a unique root given a sequence,
namely, in all other cases, the duplication root is not necessarily
unique. The characterization is stated in Theorem \ref{th:all}. Before
moving to Theorem \ref{th:all}, consider the following example:

\begin{example}\label{exm:contr}
  Let $U = \mathset{2,3,4}$ be a set of duplication lengths and $\Sigma =
  \mathset{1,2,3}$. Consider
  \[z=\lefteqn{\overbrace{\phantom{\,1\,2\,3\,2\,1\,2\,3\,2}}^{\alpha\alpha}}1\,2\,3\,2\,1\underbrace{\,2\,3\,2\,3}_{\beta\beta}.\]
  The sequence $z$ has two tandem repeats $\alpha\alpha$ and
  $\beta\beta$ with $\abs{\alpha} = 4$ and $\abs{\beta} = 2$. If we
  deduplicate $\alpha\alpha$ first from $z$ , we get
  \[ 123212323 \dedup_{4} 12323 \dedup_{2} 123.\]
  However, if we deduplicate $\beta\beta$ first from $z$ we
  get
  \[123212323 \dedup_{2} 1232123.\]
\end{example}

Theorem \ref{th:all} generalizes the statement presented in the
example above to any set of duplication lengths. We naturally extend
all previous notation to allow duplication and deduplication of
several lengths by replacing the usual $k$ subscript with a set $U$,
where $U\subseteq \N$. For example, $R_U(z)$ denotes the set of roots
obtained via a sequence of deduplications of lengths from $U$,
starting with the string $z$. The property we would like to study is
formally defined next.

\begin{definition}
  Let $\Sigma\neq\emptyset$ be an alphabet, and $U\subseteq\N$,
  $U\neq\emptyset$, a set of tandem-duplication lengths. We say
  $(\Sigma,U)$ is a \emph{unique-root pair}, iff for all
  $z\in\Sigma^*$ we have $\abs{R_U(z)}=1$. Otherwise, we call $(\Sigma,U)$
  a \emph{non-unique-root pair}.
\end{definition}

We observe that the actual identity of the letters in the alphabet is
immaterial, and only the size of $\Sigma$ matters. Additionally,
simple monotonicity is evident: If $(\Sigma,U)$ is a unique-root pair,
then so is $(\Sigma',U)$, for all $\Sigma'\subseteq
\Sigma$. Similarly, if $(\Sigma,U)$ is a non-unique-root pair, then so
is $(\Sigma',U)$, for all $\Sigma\subseteq\Sigma'$.

The following sequence of lemmas will provide the basis for a full
classification of unique-root pairs.

\begin{lemma}
  \label{lem:alpha1}
  Let $\Sigma=\mathset{a}$ be an alphabet with only a single letter.
  Let $U\subseteq\N$, and denote $k=\min(U)$. Then $(\Sigma,U)$ is a
  unique-root pair if and only if $k|m$ for all $m\in U$.
\end{lemma}
\begin{IEEEproof}
  If $k|m$ for all $m\in U$, then any sequence $a^n$, $n\in\N$ has a
  unique root
  \[ a^n\dedup^*_U a^{n\bmod k},\]
  where in the expression above $n\bmod k$ denotes the unique integer
  from $\mathset{1,2,\dots,k}$ with the same residue modulo $k$ as
  $n$.

  In the other direction, if there exists $m\in U$ such that $k\nmid m$, let
  us consider the sequence $a^{k+2m}$. By first deduplicating a length $m$
  sequence, and then as many deduplications of length $k$ we obtain
  \[ a^{k+2m} \dedup_U a^{k+m} \dedup^*_U a^{m\bmod k}=x.\]
  However, by only deduplicating length $k$ sequences, we also get
  \[ a^{k+2m} \dedup^*_U a^{2m\bmod k}=y.\]
  Both $x$ and $y$ are irreducible since $1\leq \abs{x},\abs{y}\leq
  k$. However, since $m\neq 0\pmod{k}$, we have
  \[ m\not\equiv 2m \pmod{k},\]
  and therefore $x\neq y$, and $a^{k+2m}$ has two distinct roots.
\end{IEEEproof}




\begin{lemma}
  \label{lem:kkm}
  Let $\Sigma$ be an alphabet, $\abs{\Sigma} \geq 2$, $km>1$, and $U
  = \mathset{k, k+m}\cup V$, where $V\subseteq
  \N\setminus\mathset{1,2,\dots,k+m}$. Then $(\Sigma,U)$ is a
  non-unique-root pair.
\end{lemma}
\begin{IEEEproof}
  By Lemma \ref{lem:alpha1} and monotonicity, if $k\nmid m$, then
  $(\Sigma,U)$ is already a non-unique-root pair, and we are done.
  Thus, for the rest of the proof we assume $m=\ell k$, for some
  $\ell\in\N$.
  
  Let $a,b\in\Sigma$ be two distinct letters, and let $v_1 v_2\dots
  v_{k+m}\in\Sigma^{k+m}$ be a sequence defined as follows:
  \[ v_i = \begin{cases}
    a & \text{$i< k+m$ and $\ceilenv{i/k}$ is odd,}\\
    b & \text{$i< k+m$ and $\ceilenv{i/k}$ is even,}\\
    v_{m} & \text{$i=k+m$.}
  \end{cases}
  \]
  Consider now the sequence
  \[z= v_1 v_2\dots v_{k+m} v_1 v_2\dots v_{k+m} v_{m+1}\dots v_{k+m-1}.\]
  We can write $z$ as
  \begin{align*}
    z &= (v_1 v_2\dots v_{k+m-1} v_m)^2 v_{m+1}\dots v_{k+m-1} \\
    &= v_1 v_2\dots v_{k+m-1} v_m v_1 v_2\dots v_{m-1}(v_m v_{m+1}\dots v_{k+m-1})^2.
  \end{align*}
  As is evident, there are two squares in $z$, one of which is of
  length $2k+2m$ and the other is of length $2k$. Deduplicating the
  square of length $2k+2m$ in $z$ first gives
\begin{align*}
  z &\dedup_U v_1 v_2\dots v_{k+m-1} v_m v_{m+1}\dots v_{k+m-1} \\
  &\dedup_U v_1 v_2\dots v_{k+m-1} = y.
\end{align*}
Deduplicating the square of length $2k$ first gives 
\[
z\dedup_U v_1 v_2\dots v_{k+m-1} v_m v_1 v_2\dots v_{k+m-1} = x.
\]

We note that $\abs{x} = 2k+2m-1$ and $\abs{y} = k+m-1$. Thus, if
further deduplications are possible, they must be deduplications of
length $k$, since both $x$ and $y$ are too short to allow
deduplications of other allowed lengths from $U$. We observe that $y$
is certainly irreducible, since it is made up of alternating blocks of
$a$'s and $b$'s of length $k$.  However, it is conceivable that $x$
may be further deduplicated to obtain $y$.

We recall $m=\ell k$. Depending on the parity of $\ell$, we have two
cases. If $\ell$ is even, we can write explicitly
\begin{align*}
  y &= (a^k b^k)^{\ell/2} a^{k-1},\\
  x &= (a^k b^k)^{\ell/2} a^{k-1}b (a^k b^k)^{\ell/2} a^{k-1}.
\end{align*}
The sequence $x$ may be further deduplicated, by noting the square $ba^{k-1}ba^{k-1}$, to obtain
\[ x \dedup_U (a^k b^k)^{\ell} a^{k-1} = x'.\]
We easily observe that $x'$ is irreducible, and $x'\neq y$ since their
lengths differ, $\abs{y}=(\ell+1)k-1$, $\abs{x}=(2\ell+1)k-1$, and
$\ell\geq 1$.

If $\ell$ is odd, we explicitly write
\begin{align*}
  y &= (a^k b^k)^{(\ell-1)/2} a^k b^{k-1},\\
  x &= (a^k b^k)^{(\ell-1)/2} a^k b^{k-1} a (a^k b^k)^{(\ell-1)/2} a^k b^{k-1}.
\end{align*}
We recall our requirement that $km>1$, which translates to $k\geq 1$,
$\ell\geq 1$ and odd, but not $k=\ell=1$. If $k\geq 2$ and $\ell\geq
3$, we easily see that $x$ is irreducible, $x\neq y$. If $\ell=1$ and
$k\geq 2$, we have $x = a^k b^{k-1} a^{k+1} b^{k-1}$ which is again irreducible,
and $x\neq y$. The final case is $k=1$ and $\ell\geq 3$, in which
\[ x = (ab)^{(\ell-1)/2} a^2 (ab)^{(\ell-1)/2} a \dedup_U^* (ab)^{\ell-1} a = x'\]
by twice deduplicating the square $a^2$. However,
$y=(ab)^{(\ell-1)/2}a$, and $y\neq x$ since $\abs{y}=1+(\ell-1)/2$ and
  $\abs{x}=\ell$, while $\ell\geq 3$.
\end{IEEEproof}

\begin{lemma}\label{lem:kk12}
For any alphabet $\Sigma$, $\abs{\Sigma} \geq 3$, and for any
$V\subseteq\N\setminus\mathset{1,2,3}$, $V\neq\emptyset$, if $U =
\mathset{1,2} \cup V$, then $(\Sigma,U)$ is a non-unique-root pair.
\end{lemma}
\begin{IEEEproof}
Let $a, b, c\in\Sigma$ be distinct symbols, and let $m = \min(V)$.
Consider the sequence
\[z = ab^{m-3}caab^{m-3}ca.\]
We now have the following two distinct roots,
\begin{align*}
  z & \dedup_{U} ab^{m-3}ca \dedup_U^* abca, \\
  z & \dedup_U ab^{m-3}cab^{m-3}ca \dedup_U^* abcabca.
\end{align*}
\end{IEEEproof}

\begin{lemma}\label{lem:kk123}
For any alphabet $\Sigma$, $\abs{\Sigma} \geq 3$, and for any
$V\subseteq\N\setminus\mathset{1,2,3}$, $V\neq\emptyset$, if $U =
\mathset{1,2,3} \cup V$, then $(\Sigma,U)$ is a non-unique-root pair.
\end{lemma}
\begin{IEEEproof}
  Let $a, b, c\in\Sigma$ be $3$ distinct symbols. Consider the sequence
  \[z = ab^{m-3}cbab^{m-3}cbc,\]
  where $m = \min(V)$. We now have the following two distinct roots,
  \begin{align*}
    z & \dedup_U ab^{m-3}cbc \dedup_U^* abcbc \dedup_U abc, \\
    z & \dedup_U ab^{m-3}cbab^{m-3}c \dedup_U^* abcbabc.
  \end{align*}
\end{IEEEproof}

We are now in a position to provide a full classification of
unique-root pairs.

\begin{theorem}
  \label{th:all}
  Let $\Sigma\neq\emptyset$ be an alphabet, and $U\subseteq\N$,
  $U\neq\emptyset$, a set of tandem-duplication lengths. Denote
  $k=\min(U)$. Then $(\Sigma,U)$ is a unique-root pair if and only if
  it matches one of the following cases:
  \begin{center}
    \begin{tabular}{c|l}
      \hline
      $\abs{\Sigma}=1$ & $U\subseteq k\N$ \\ \hline
      \multirow{2}{*}{$\abs{\Sigma}=2$} & $U=\mathset{k}$ \\
      & $U\supseteq \mathset{1,2}$ \\ \hline
      \multirow{3}{*}{$\abs{\Sigma}\geq 3$} & $U=\mathset{k}$ \\
      & $U=\mathset{1,2}$ \\
      & $U=\mathset{1,2,3}$ \\ \hline
    \end{tabular}
  \end{center}
\end{theorem}

\begin{IEEEproof}
  The case of $\abs{\Sigma}=1$ is given by Lemma \ref{lem:alpha1}. The
  case of $\abs{U}=1$ was proved in \cite{LeuMarMit05}, with an
  alternative proof we provided in Section \ref{sec:k-tandem}. The
  case of $\abs{\Sigma}=2$ and $\mathset{1,2}\not\subseteq U$, was
  proved in Lemma \ref{lem:kkm}. It is also folklore that having
  $\abs{\Sigma}=2$ and $\mathset{1,2}\subseteq U$ gives a unique-root
  pair, since we can always deduplicate runs of symbols to single
  letters, and then deduplicate pairs, to obtain one of only six
  possible roots: $a$, $b$, $ab$, $ba$, $aba$, $bab$. The choice of
  root depends only on the first letter of the word, its last letter,
  and when they're the same, on the existence of a different letter
  inside. All deduplication actions do not change those, regardless of
  the length of the deduplication.

  When $\abs{\Sigma}\geq 3$, the unique-root property for
  $U=\mathset{1,2}$ and $U=\mathset{1,2,3}$ was established in
  Corollary \ref{cor:unqroot2} and Theorem \ref{th:unqroot3},
  respectively. The non-unique-root property for the other cases was
  proved in Lemma \ref{lem:kkm}, Lemma \ref{lem:kk12}, and Lemma
  \ref{lem:kk123}.
\end{IEEEproof}

\section{Conclusion}
\label{sec:conc}

We provided error-correcting codes, and in some cases, exact capacity,
for the tandem-duplication channel. These codes mostly rely on
unique-root pairs of alphabets and duplication lengths. Several
interesting questions remain open. In particular, we do not know yet
how to construct general $(n,M;*)_U$ over $\Sigma$, especially when we
do not necessarily have unique roots. We also mention the interesting
combinatorial problem of counting the number of distinct roots of a
string, and finding strings of a given length with as many roots as
possible.

\bibliographystyle{IEEEtranS}
\bibliography{allbib}

\appendix

We provide a short proof of \eqref{eq:cap0k}. We need to estimate the
largest eigenvalue of $A_q(k)$ from \eqref{eq:mat0k}, i.e., to estimate
the largest root $\lambda$ of its characteristic polynomial
\[ \chi_{A_q(k)}(x) = \frac{x^{k+2}-qx^{k+1}+q-1}{x-1}.\]
Since this largest root is strictly greater than $1$, we can alternatively
find the largest root of the polynomial
\[ f(x) = x^{k+2}-qx^{k+1}+q-1.\]

We shall require the following simple bounds. Taking the first term in
the Taylor expansion of $e^x$, and the error term, we have for all
$x>0$,
\[ e^x = 1 + x e^{x'},\]
for some $x'\in [0,x]$. Since $x>0$ and $e^x$ is increasing, we have
\[ e^x = 1 + x e^{x'} \leq 1 + x e^x,\]
or alternatively,
\begin{equation}
  \label{eq:ex}
  1 - e^x \geq -xe^x.
\end{equation}
Similarly, taking the first two terms of the Taylor expansion, for all
$x>0$, we get the well known
\begin{equation}
  \label{eq:exx}
  e^x > 1+x.
\end{equation}

We return to the main proof. In the first direction, let us first
examine what happens when we set
\[ x=q e^{-\frac{q-1}{q^{k+2}}}.\]
Then
\begin{align*}
  f(x) &= q^{k+2} e^{-\frac{q-1}{q^{k+2}}(k+2)} - q^{k+2} e^{-\frac{q-1}{q^{k+2}}(k+1)}
  +q-1 \\
  &= q^{k+2} e^{-\frac{q-1}{q^{k+2}}(k+2)}\parenv{1-e^{\frac{q-1}{q^{k+2}}}}
  +q-1 \\
  &\overset{\text{(a)}}{\geq} (q-1) \parenv{1- e^{-\frac{q-1}{q^{k+2}}(k+1)}} \\
  &> 0,
\end{align*}
where (a) follows by an application of \eqref{eq:ex}.

In the other direction, we examine the value of $f(x)$ when we set
\[ x=q e^{-\frac{q-1}{q^{k+2}}\alpha},\]
where $\alpha$ is a constant depending on $q$ and $k$. To specify
$\alpha$ we recall $W(z)$, $z\geq -\frac{1}{e}$, denotes the Lambert
$W$-function, defined by
\[ W(z)e^{W(z)}=z.\]
We define
\[ \alpha = \frac{W\parenv{ -\frac{q-1}{q^{k+2}}(k+2) }}{-\frac{q-1}{q^{k+2}}(k+2)} = e^{-W\parenv{ -\frac{q-1}{q^{k+2}}(k+2) }}.\]
Except for $k=1$ and $q=2$, for all other values of the parameters we
have
\[-\frac{q-1}{q^{k+2}}(k+2) \geq -\frac{1}{e},\]
rendering the use of the $W$ function valid.
We also note that for these parameters we have $\alpha\geq 1$.

Let us calculate $f(x)$,
\begin{align*}
  f(x) &= q^{k+2} e^{-\frac{q-1}{q^{k+2}}(k+2)\alpha} - q^{k+2} e^{-\frac{q-1}{q^{k+2}}(k+1)\alpha}
  +q-1 \\
  &= q^{k+2} e^{-\frac{q-1}{q^{k+2}}(k+2)\alpha}\parenv{1-e^{\frac{q-1}{q^{k+2}}\alpha}}
  +q-1 \\
  &\overset{\text{(a)}}{<} (q-1) \parenv{1- \alpha e^{-\frac{q-1}{q^{k+2}}(k+2)\alpha}}\\
  &\overset{\text{(b)}}{=} (q-1)\parenv{1-1} = 0,
\end{align*}
where (a) follows by an application of \eqref{eq:exx}, and (b) follows
by substituting the value of $\alpha$.

In summary, $f(x)$ is easily seen to be decreasing in the range
$[1,(k+1)q/(k+2)]$, and increasing in the range $[(k+1)q/(k+2),\infty)$,
and therefore, its unique largest root $\lambda$ is in the range
\[ q e^{-\frac{q-1}{q^{k+2}}\alpha} \leq \lambda \leq q e^{-\frac{q-1}{q^{k+2}}}.\]
It is easy to verify that $\alpha=1+o(1)$, where $o(1)$ denotes a function
decaying to $0$ as $k\to\infty$. Hence,
\[ \lambda = q e^{-\frac{q-1}{q^{k+2}}(1+o(1))},\]
and therefore
\begin{align*}
  \ccap(\rll_q(0,k)) &= \log_2 \lambda \\
  &= \log_2 q - \frac{(q-1)\log_2 e}{q^{k+2}}(1+o(1)).
\end{align*}

\end{document}